\documentclass[aps,pre,nofootinbib,showpacs,twocolumn,10pt]{revtex4-1}%
\usepackage{amssymb,latexsym,mathrsfs}
\usepackage{amsfonts}
\usepackage{hyperref}
\usepackage{mathrsfs}
\usepackage{amsmath}
\usepackage{graphicx}
\usepackage{color}

\newcommand{\beq}{\begin{equation}}
\newcommand{\eeq}{\end{equation}}
\newcommand{\beqn}{\begin{eqnarray}}
\newcommand{\eeqn}{\end{eqnarray}}
\newcommand{\bearr}{\begin{array}}
\newcommand{\enarr}{\end{array}}

\def\bea{\begin{eqnarray}}
\def\eea{\end{eqnarray}}
\def\ba{\begin{array}}
\def\ea{\end{array}}

\begin{document}
\title{Thermally driven classical Heisenberg model in 1D with a local time varying field} 
\author{Debarshee  Bagchi}
\email[E-mail address: ]{debarshee.bagchi@saha.ac.in}
\affiliation{Theoretical Condensed Matter Physics Division,\\ Saha Institute of Nuclear Physics,
Kolkata, India.}
\date{\today}

\begin{abstract}
We study thermal transport in the one dimensional classical Heisenberg
model driven by boundary heat baths in presence of a local time varying
magnetic field that acts at one end of the system. The system is studied
numerically using an energy conserving discrete-time odd even dynamics.
We find that the steady state energy current shows thermal resonance as the 
frequency of the time-periodic forcing is varied. When the amplitude of the
forcing field is increased the system exhibits multiple resonance peaks
instead of a single peak. Both single and multiresonance survive in the
thermodynamic limit and their magnitudes increase as the average temperature
of the system is decreased. Finally we show that, although a reversed thermal
current can be made to flow through the bulk for a certain range of the forcing
frequency, the system fails to behave as a heat pump, thus revalidating the
fact that thermal pumping is generically absent in such force-driven lattices.

\end{abstract}
\maketitle

\section{Introduction}

Low dimensional thermal transport has been a topic of intense research
interest in recent times. The reason for the sudden upsurge in this
field is primarily two-fold. Firstly, low dimensional systems have rich
and intriguing transport properties and therefore contribute immensely in
widening our theoretical understanding of the fundamental principles
of transport, e.g., the necessary and sufficient criteria for the validity
of Fourier law in thermally driven low dimensional systems \cite{Lebo, Lepri, AD-rev}.
Although substantial progress has been made in the theoretical
front, a comprehensive understanding is still lacking. The advancement
in low dimension experiments has also greatly motivated theoretical
research since theoretical predictions can often be readily tested in
laboratories nowadays. 
Secondly, studying thermal transport in low dimension is of
huge technological interest because of the recent breakthrough
in nanoscale thermal devices which rely on the energy transport 
properties of phonons present in a system \cite{phononics}.
Similar to its electrical counterpart, theoretical designs for
phononic devices such as thermal diodes, transistors, 
logic gates, memory devices, phonon waveguide \cite{dv1, dv2, dv3, dv4, dv5} 
have been proposed and some of them also experimentally realized recently. 
Thus, it has been possible to control and manipulate heat current, 
just as one would control electrical current in electronic devices \cite{control}. 
These new exciting developments have induced a flurry of active
theoretical and experimental research in this field at present.

A particular lattice model that has been extensively studied in regards to its thermal
transport properties is the Frenkel-Kontorova (FK) model. The FK model is a nonlinear
lattice model which exhibits normal transport of thermal energy in one dimension (1D)
and thus obeys the Fourier law \cite{FK-r}. The FK model is particularly significant because
of the fact that theoretical designs of many phononic devices have been
based on this lattice model \cite{phononics}.
It has been shown in a recent paper \cite{reso} that the thermally driven FK model
in presence of local time-periodic forcing at one end of the system shows thermal
resonance - the energy current through the system attains a maximum value for
some characteristic frequency of the external forcing.
It was also suggested that a thermal pump can be designed using this setup which can
direct thermal energy against the imposed gradient. 
Subsequently it was shown \cite{multi} that the thermal resonance seen in
this model is actually a multiresonance phenomenon which appears in certain
ranges of the system parameter. It was also argued (based on direction of energy
flow in the system) that thermal pumping is actually absent in such force-driven lattices.
Other nonlinear models with local time-periodic driving at one edge of the system
have also been studied 
recently, such as, the Fermi-Pasta-Ulam model, sine-Gordon model, Klein-Gordon
model, discrete nonlinear Schr\"{o}dinger system \cite{nlm1,nlm2,nlm3,nlm4,nlm5}.

Motivated by these results, we investigate the transport properties of the thermally
driven classical Heisenberg model in presence of local time varying forcing. 
The classical Heisenberg model \cite{fisher,joyce}
has been studied, both analytically and numerically, for several decades and has
become a prototypical model for magnetic insulators. Although the transport properties
of its quantum counterpart (spin-$\frac12$ quantum Heisenberg model) have been extensively 
studied \cite{QM1,QM2,QM3}, not much attention has been paid to the thermal transport
properties of the classical Heisenberg model until recently \cite{savin,huse,our,savin-sp}.
The thermal transport properties of the classical model is found to obey the Fourier
law and this diffusive energy transport is attributed to the nonlinear spin wave
interactions of the system \cite{savin}.
A deeper understanding of thermal properties of such spin systems is extremely 
desirable now since 
several new magnonic devices, such as, memory elements, logic gates, switches,
waveguides etc. are now being conceived by manipulating spin waves by external
magnetic field in ferromagnetic materials \cite{magnonics1, magnonics2}.
Transport in spin systems can also give rise to intriguing phenomenon
such as spin Seebeck effect where a spin voltage is generated by a
temperature gradient even across a ferromagnetic insulator and
is due to the thermally excited spin wave interactions about
which very little is known \cite{sse1,sse2}.

In this paper, we study the classical Heisenberg model in presence of a time varying
magnetic field that acts locally at one end of the one dimensional system and a
thermal bias is imposed by heat baths attached at the two boundaries. Eventually
this force-driven system attains an oscillatory nonequilibrium steady state. In
the steady state, the thermal current flowing through the system shows thermal
resonance at a characteristic frequency of the time-periodic force, similar to the
FK model. 
We discuss the dependence of system parameters on the resonance and find that the
Heisenberg model is quite similar to FK model (which, in turn, has resonance
properties  similar to the harmonic lattice \cite{multi}) in regards to the resonance both
exhibit, although the mechanism of energy transport in the FK model and the
spin system are quite different.
We also investigate the multiresonance feature that can be observed in this model
in some parameter range and study its parameter dependence as well. We point out the
typical features of the multiresonance seen in our model and compare it with that of the
FK/harmonic system.
Finally, we explicitly demonstrate that, although the direction of the current in the bulk
can be reversed when the system resonates, one can not use this as a thermal pump.

The remainder of the paper is organised as follows. We define the classical
Heisenberg model in detail and discuss our numerical scheme in Sec. \ref{model} briefly.
Thereafter in Sec. \ref{results}, we present our numerical results and
demonstrate the existence of thermal resonance in this system. 
We study dependencies of system parameters on the resonance
and show that the occurrence a multiresonance phenomenon for certain
parameter ranges. We then demonstrate the absence of thermal pumping in this
system which reconfirms the fact that thermal pumping is generically absent in
such force-driven lattices. 
Finally, we conclude by summarizing our main results in Sec. \ref{summary}.

\section{Model and Numerical scheme}
\label{model}
\begin{figure}[htbp]
\includegraphics[width=8cm,height=3.5cm,angle=0]{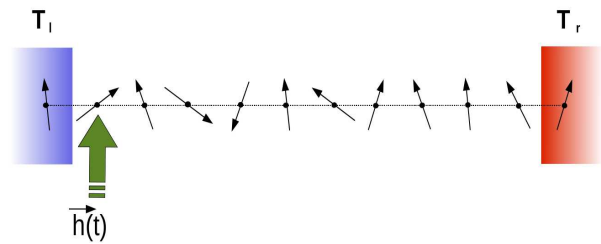}\hfill
\caption{(Color online) Schematic diagram of the classical Heisenberg model in one dimension
with two boundary heat baths at temperatures $T_l$ and $T_r$, and local time varying 
magnetic field $\vec h(t)$ that acts on the first site of the system $i = 1$.
We choose $T_l < T_r$ and $\vec h(t) \equiv (0,0,A \sin(\omega t))$.}
\label{fig:schematic}
\end{figure}
Consider classical Heisenberg spins $\{\vec{S_i}\}$ (three-dimensional unit
vectors) on a one-dimensional regular lattice of length $N$ $(1 \le i \le N)$
and the Hamiltonian of the system is given as,
\begin{equation}
\mathcal{H} = -K \sum_{i=1}^{N}\vec{ S}_i\cdot \vec{ S}_{i+1} - \vec h(t) \cdot \vec S_i \delta_{i1},
\label{ham}
\end{equation}
where the spin-spin interaction is ferromagnetic for coupling 
$K > 0$ and anti-ferromagnetic for $K < 0$. 
The second term Eq. (\ref{ham}) is due to a time varying magnetic field that
acts locally on the spin at the first site $i=1$.
A schematic diagram of our system is shown in Fig. \ref{fig:schematic}.
The microscopic equation of motion for the spin vectors can be written as
\begin{equation}
\frac d{dt} {\vec{ S}_i} = \vec{S}_i \times \left[\vec{B}_i + \vec h(t) \delta_{i1}\right],
\label{eom}
\end{equation}
where $\vec{B}_i = K(\vec{S}_{i-1} + \vec{S}_{i+1})$ is the local molecular field
experienced by the spin at site $i$. Apart from the local forcing, there is also an
overall thermal gradient across the system maintained by two heat baths attached
at the two boundaries. This is implemented by introducing to additional spins
at sites $i = 0$ and $i = N+1$. The bonds between the pairs of spins ($\vec{S}_0, \vec{S}_1$) and 
($\vec{S}_N, \vec{S}_{N+1}$) at two opposite ends of the system behave as 
stochastic thermal baths. The left and right baths are in equilibrium at 
their respective temperatures, $T_l$ and $T_r$ and the bond energies
$E_l = -K \vec S_0 \cdot  \vec S_1$ and $ E_r = -K \vec S_N \cdot  \vec S_{N+1}$ 
have a Boltzmannian distribution.
%
%
%
%
The interaction strength of the bath spins with the system is taken to 
be $K$, and therefore both $E_l$ and $E_r$ are bounded in the range $(-K, K)$.
Thus the average energies of the baths are given by
$ \langle E_{l}\rangle  = - K \mathcal{L}(K/T_l) ~ {\rm and}~  \langle E_{r} \rangle 
= - K \mathcal{L}(K/T_r)$, $\mathcal{L}(x)$ being the standard Langevin function.
Thus one can set the two baths at a fixed average energy (or temperature)
and a thermal current flows through the system if $T_l \neq T_r$. For our
simulations we choose $T_l < T_r$ and therefore the periodic forcing is
near the colder boundary.

We investigate the steady state transport properties of the Heisenberg model
by numerically computing quantities, such as, currents, 
energy profiles etc. using the DTOE dynamics \cite{our}. 
Briefly stated, the DTOE dynamics updates the spins belonging to the
odd and even sub-lattices alternately using a spin precessional dynamics
\begin{equation}
\vec{S}_{i,t+1} = \left[\vec{S} \cos \phi + (\vec{S} \times \hat{B}) \sin \phi + (\vec{S}\cdotp\hat{B})\hat{B}(1 - \cos \phi) \right]_{i,t},
\label{precess}
\end{equation}
where $\hat{B}_i = \vec{B}_i/|\vec{B}_i|$, $\phi_i = |\vec{B}_i| \Delta t$ and $\Delta t$
is the integration time step. The above formula is also sometimes referred to as the
\textit{rotation formula} and holds for rotation of any finite magnitude \cite{goldstein}.
Note that $\vec B_i$ in the above equation has to be replaced by the vector
($\vec B_i + \vec h(t))$ for the first site $i = 1$.
%
%
The advantages of using the DTOE dynamics are as follows. This dynamics is strictly
energy conserving and maintains the length of the spin vectors naturally.
The total spin conservation is also approximately maintained and is comparatively
better than other conventional integration schemes \cite{sdiff}. Also it has been
thoroughly verified that local thermal equilibrium is established in the thermally
driven system (in absence of forcing) when evolved using the DTOE dynamics
independently of the value of the integration time step $\Delta t$.
A thorough discussion of the DTOE scheme and numerical implementation
of the thermal baths have been presented in Ref. \cite{our}.

The computation of the steady state thermal current is done as described in the following.
Since the DTOE dynamics alternately updates only half of the spins (odd/even)
but all the bond energies simultaneously, the energy of the bonds $E^o_i$
measured immediately after the update of odd spins is different from 
$E^e_i$ measured after the update of even spins where 
$E_i = -K \vec{ S}_i\cdot \vec{ S}_{i+1} - \vec h(t) \cdot \vec S_i \delta_{i1}$.
Clearly, the difference $E^e_i - E^o_i$ is a measure of the energy
flowing through the $i$-th bond in time $\Delta t$. Thus the thermal
current (rate of flow of thermal energy) in the steady state is given by
\begin{equation}
j = \langle E^e_i - E^o_i  \rangle/\Delta t
\label{J}
\end{equation}
and the average energy of $i$-th bond is $E_i = \frac 12 \langle E^e_i + E^o_i \rangle$.
As already mentioned, the thermal current $j$ in the one dimensional classical Heisenberg model obeys
the Fourier law, i.e., $j \sim 1/N$. Let us define a total current $J \equiv j N$
which clearly is independent of the system size. In the following, we present our numerical
results in terms of the total current $J$, since this will allow us to compare
the thermal current of systems of different sizes.

\section{Numerical Results}
\label{results}

\subsection{Resonance}
For simulations, we choose the time varying magnetic field of the form
$\vec h(t) \equiv (0,0,A \sin(\omega t))$ and the boundary heat baths
have average energy $\langle E_l \rangle = E_b$ and 
$\langle E_r \rangle = E_b + \Delta E$. Starting from a random
initial configuration of spins, the system is evolved using the DTOE
dynamics with integration time-step $\Delta t = 0.25$. The spin system
is first relaxed, typically for $\sim 10^6-10^7$ iterations, and thereafter the
steady state quantities are computed for the next $\sim 10^7-10^8$ iterations
which is also averaged over several independent realizations (typically $\sim 1000$)
of the initial spin configuration. In absence of periodic forcing, a
current flows through the system in response to the imposed
thermal bias. To fix the sign of the thermal current we set the following
convention - a current flowing from a larger to a smaller site index $i$
is negative; in other words, a current from the right towards the left
end of the lattice is taken to be negative and vice-versa.

\begin{figure}[htbp]
\includegraphics[width=6.5cm,angle=-90]{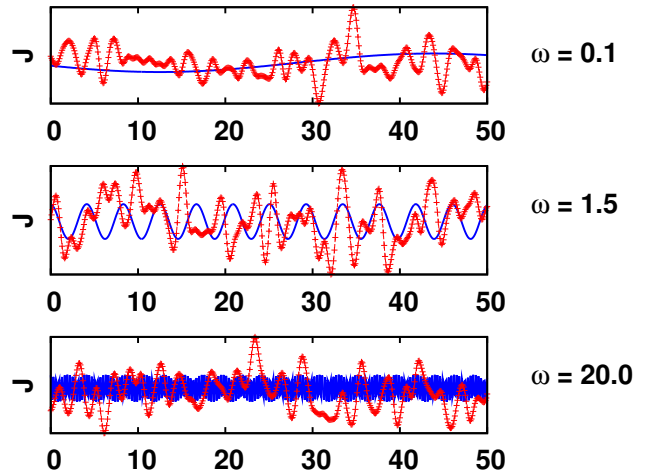}
\caption{(Color online) Typical oscillations of the total thermal current $J$ 
(data points shown as `+' symbol) in the steady state at small ($\omega = 0.1$), 
moderate ($\omega = 1.5$) and high  ($\omega = 20.0$) frequencies along with
the corresponding time varying magnetic field (smooth continuous lines). 
Thermal resonance occurs in our model at frequency $\omega \approx 1.5$
where the typical oscillation frequency of the current becomes comparable
to the periodic forcing frequency.}
\label{fig:ts}
\end{figure}
\begin{figure}[htbp]
\hskip-1.1cm
\includegraphics[width=6.65cm,angle=-90]{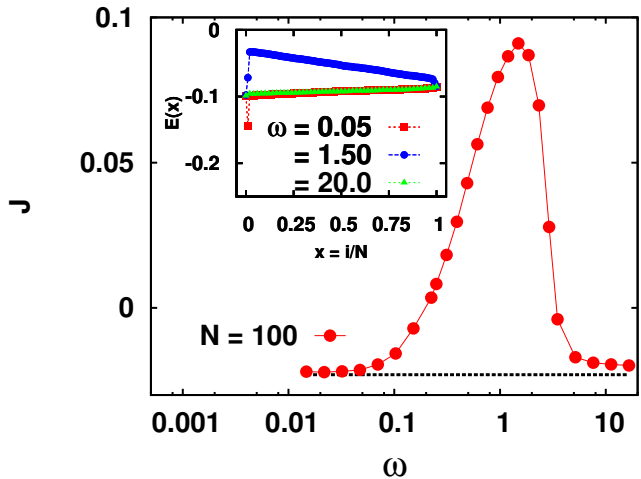}
\caption{(Color online) Thermal resonance in classical Heisenberg model: (a) thermal current
$J$ shows thermal resonance for some value of the forcing frequency $\omega \approx 1.5$
where the $J(\omega)$ curve shows a peak. The low and high frequency behavior is 
same as that in absence of any periodic forcing (shown as a dotted line); The parameters
used are $N = 100, A = 1.0, E_b = -0.1$, and $\Delta E = 0.015$.
(Inset) The change of sign of the slope of the energy profile for $\omega = 1.5$ as compared to
$\omega = 0.05, 20.0$ suggests current reversal in the bulk of the system.
}
\label{fig:reso100}
\end{figure}

We categorize the frequency response of the system in three separate regimes - the
low, moderate and high frequency. For comparison, we plot the thermal current $J$ and
the periodic forcing as a function of time, for three frequencies corresponding to the
three regimes mentioned above. 
This is shown in Fig. \ref{fig:ts} where we have shown the thermal current and
the sinusoidal forcing field for frequencies $\omega = 0.1, 1.5, 20.0$.
The steady state current as a function of the forcing frequency is shown in 
Fig. \ref{fig:reso100}. We find that the thermal transport scenario is
essentially the same for the small and high frequency regimes.
For small values of the forcing frequency the current flows through the system from 
the higher temperature end to the lower temperature end (hence negative according
to our convention) and similarly for high frequencies, and the average value of
the current is roughly the same in these to regimes. This can be understood
from the fact that for very small frequencies the typical timescales of the system
are much smaller than the forcing timescale and thus the system senses two opposite
static forces which amounts to no net forcing. In the other limit,
the system fails to respond to the rapidly varying forcing and thus effectively
experiences no external forcing. Thus in the two asymptotic limits the frequency
behavior is essentially the same as can be clearly seen from Fig. \ref{fig:reso100}.

\begin{figure}[htbp]
\centering\hskip-0.85cm
\includegraphics[width=6.5cm,angle=-90]{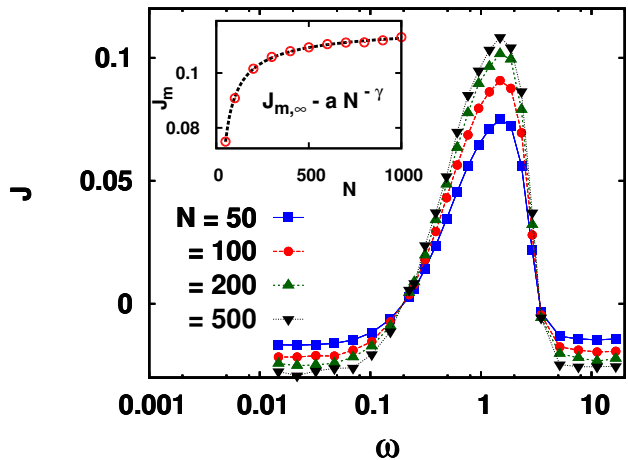}
\caption{(Color online) Thermal resonance survives in the thermodynamic limit as can understood from the
$J(\omega)$ curve for system sizes $N = 50, 100$ and $200$, all other parameters remaining the same
as Fig. \ref{fig:reso100}. (Inset) The maximum current $J_m$ corresponding to $\omega = \omega_m$
saturates in the thermodynamic limit. The $J_m \sim N$ data obtained from simulation fits with
the form $J_{m,\infty} - a N^{-\gamma}$ (shown as a broken line) where $J_{m,\infty} = 0.118$,
$a = 0.633$, and $\gamma = 0.685$.
}
\label{fig:reso}
\end{figure}

In the moderate frequency range, where the typical oscillations of the thermal current 
is comparable to that of the external periodic forcing, the current however shows 
thermal resonance - the current attains a maximum value $J_m$ corresponding to some
characteristic resonance frequency $\omega_m$. By properly choosing the bath
temperatures, the current $J$ can even be made to change sign (here, from negative 
to a positive value). This implies that for certain range of the forcing frequency
the thermal current through the bulk of the system flows in the opposite direction
i.e., from the colder end to the hotter end. We refer to this frequency range for
which the bulk current is reversed as the resonance region for convenience sake.

\begin{figure}[htbp]
\centering\hskip-1cm
\includegraphics[width=6.5cm,angle=-90]{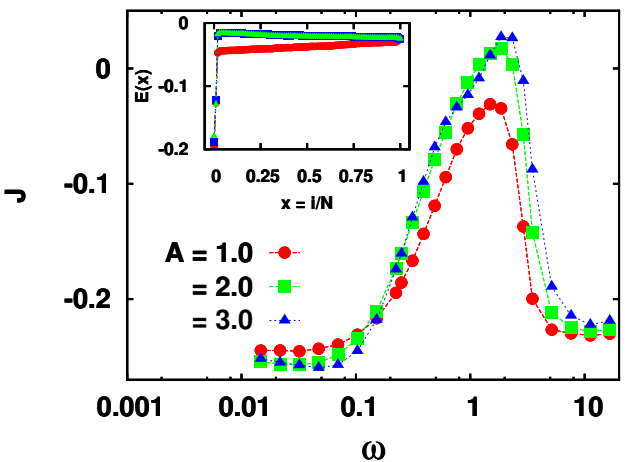}
\caption{(Color online) (a) Thermal resonance with a larger energy gradient ($E_b = -0.2$ and $\Delta E = 0.175$) 
for different values of the forcing amplitude $A$. Increasing the amplitude increases the current and makes $J > 0$;
(Inset) Energy profiles corresponding to $A = 1.0, 2.0$ and $3.0$ for $N = 100$. 
The energy profile for $A = 1.0$ has slope of opposite sign as compared to that
for $A = 2.0$ and $3.0$. 
}
\label{fig:large}
\end{figure}

To check that the thermal current indeed gets reversed for frequencies 
in the resonance region, we compute the energy profile of the system 
[see Fig. \ref{fig:reso100} (inset)]
for three forcing frequencies belonging to the three frequency regimes. 
It is found that the energy profiles for small and large frequencies have the usual
linear form connecting the two heat baths with a discontinuity at the forcing
site. The energy profile for frequency in the resonance region $\omega = 1.5$, 
however, has an opposite
slope in the bulk of the system. Thus by merely tuning the forcing frequency,
one can easily manipulate the magnitude as well as the direction of flow of
the thermal current in the bulk of the system.
Two more characteristic features of the observed thermal resonance
are in order. Firstly, the resonance effect survives in the thermodynamic
limit which is evident from Fig. \ref{fig:reso}, 
where we have shown resonance for three different system sizes. The maximum
current $J_m$ as a function of the system size fits nicely with the functional
form $J_m = J_{m,\infty} - a N^{-\gamma}$, where $J_{m,\infty}$ is the
saturation value of maximum current in the thermodynamic limit, and $a,\gamma$
are fitting parameters. Thus the maximum current has a finite limiting value
for a thermodynamically large system. This evidently shows that thermal resonance
is an intrinsic feature of the system and not a finite size effect. Secondly, the
resonance frequency $\omega_m$ seems to be completely independent of the system
size, which again points to the fact that the $\omega_m$ is an intrinsic frequency
of the system.

If the temperature gradient is comparatively large (keeping all other
parameters unchanged), the resonance phenomenon is still there but the current
through the system has now a large negative value and tuning only the frequency
can not push the current to a positive value. As before, the energy profiles can
be computed which also validate the fact that the current although is amplified
(the slope of the energy profile increases) but does not get reversed (the slope
does not change sign). This is shown in Fig. \ref{fig:large}. However, increasing
the amplitude can push the current to a positive value and current reversal
can be achieved in the bulk of the system.

\begin{figure}[htbp]
\hskip-1cm
\includegraphics[width=6.5cm,angle=-90]{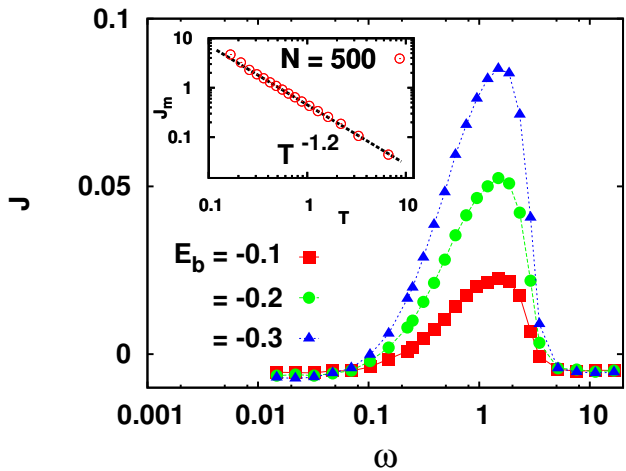}\hfill
\caption{(Color online) Thermal current with frequency for different
average energy of the bath $E_b = -0.1,-0.2$ and $-0.3$ and same energy gradient
$\Delta E = 0.015$. Resonance enhances as the temperature decreases.
(Inset) Log-log plot of the maximum thermal current $J_m$ corresponding to 
$\omega = 1.5$ for $N = 500$.
}
\label{fig:temp}
\end{figure}

We also investigate the effect of the temperature on the resonance feature.
Let the temperatures of the two baths be $T_l = T$ and $T_r = T+\Delta T$.
For the same $\Delta T$ we find that the resonance effect is enhanced at a
lower temperature and the current amplification is larger as is shown in
Fig. \ref{fig:temp}. To see how the maximum thermal current $J_m$ varies
with temperature, we set $\omega = \omega_m \approx 1.5$ and measure $J_m$
for different bath temperatures keeping the same temperature difference.
The inset in Fig. \ref{fig:temp} suggests that the maximum current $J_m$
is inversely proportional (approximately) to the temperature of the system.
Phenomenologically, the total thermal current in a finite system can be
related to its size via the relation \cite{our}
\begin{equation}
J = \kappa \frac{N \Delta E}{N + \xi},
\end{equation}
where $\kappa$ is the thermal conductivity and $\xi$ is the correlation length of
the system, both being a function of the temperature. Since both $\kappa$ and 
$\xi$ vary as $T^{-1}$ at low temperatures \cite{our}, the above relation 
suggests that, in the thermodynamically large system ($N \gg \xi$), $J_m \sim T^{-1}$.
Thus our numerical result is found to be close to this analytical result but needs
corrections for finite size.

\subsection{Multiresonance}
\begin{figure}[htbp]
\centering\hskip-1.05cm
\includegraphics[width=5.1cm,angle=-90]{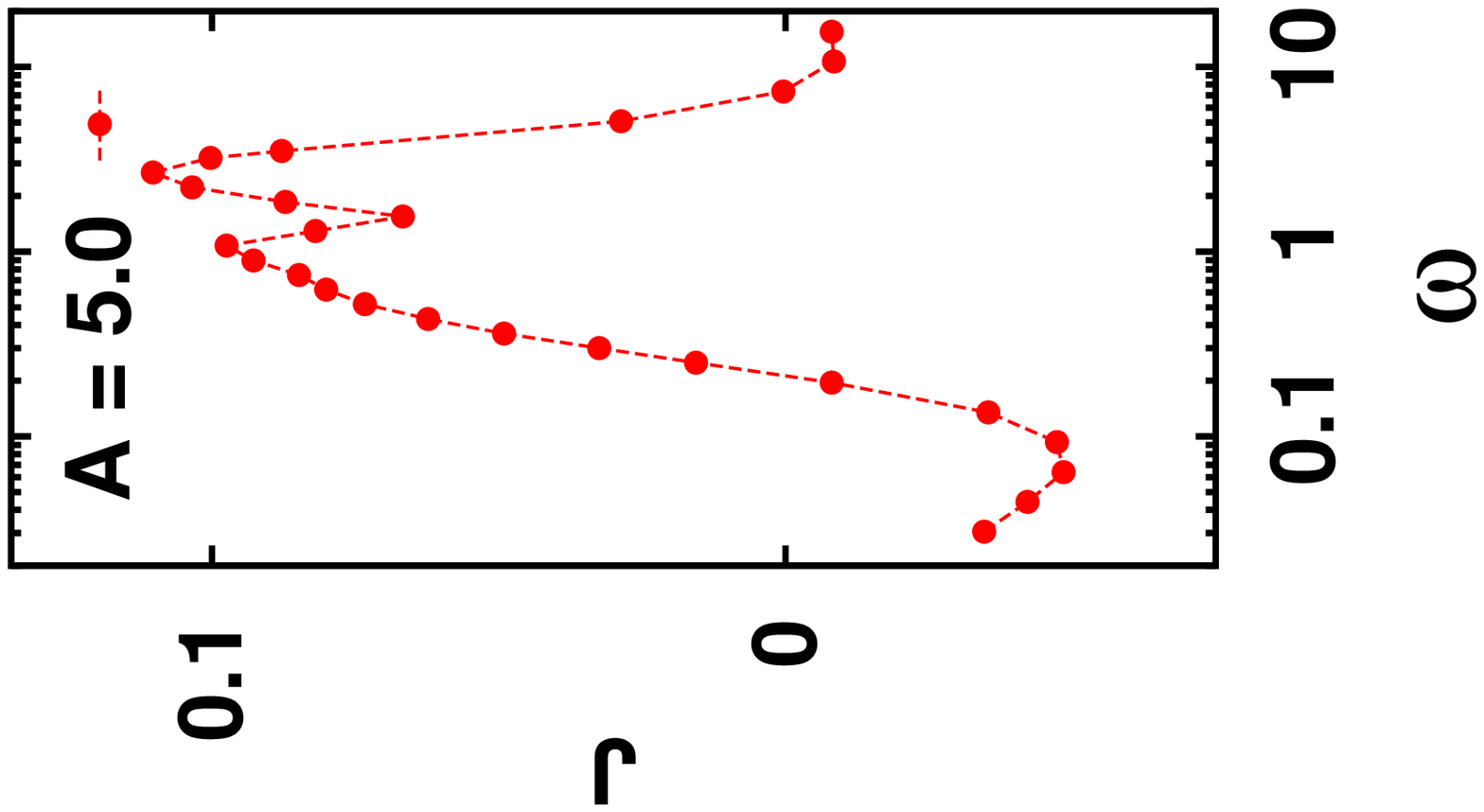}\hskip-0.8cm
\includegraphics[width=5.1cm,angle=-90]{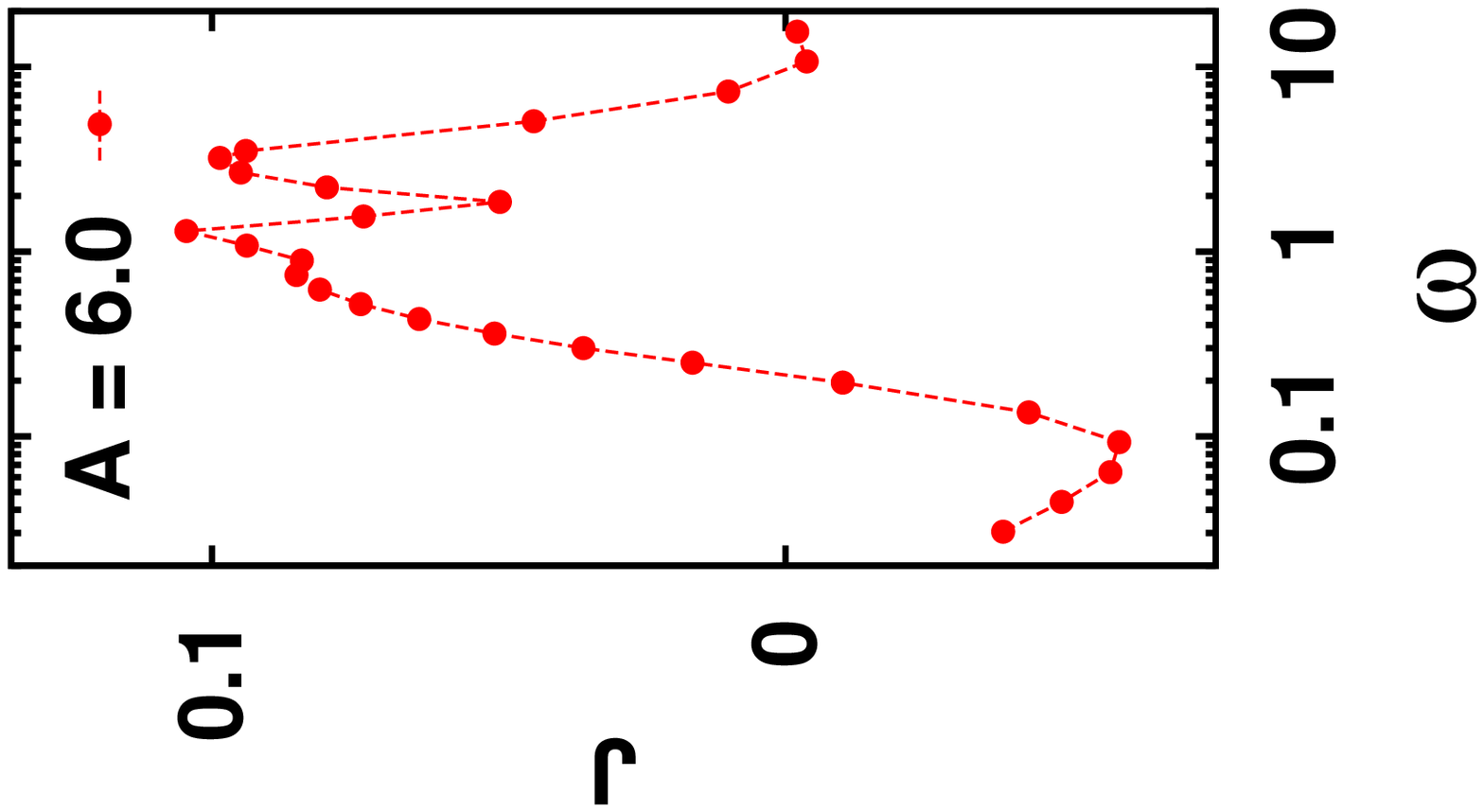}\hskip-0.8cm
\includegraphics[width=5.1cm,angle=-90]{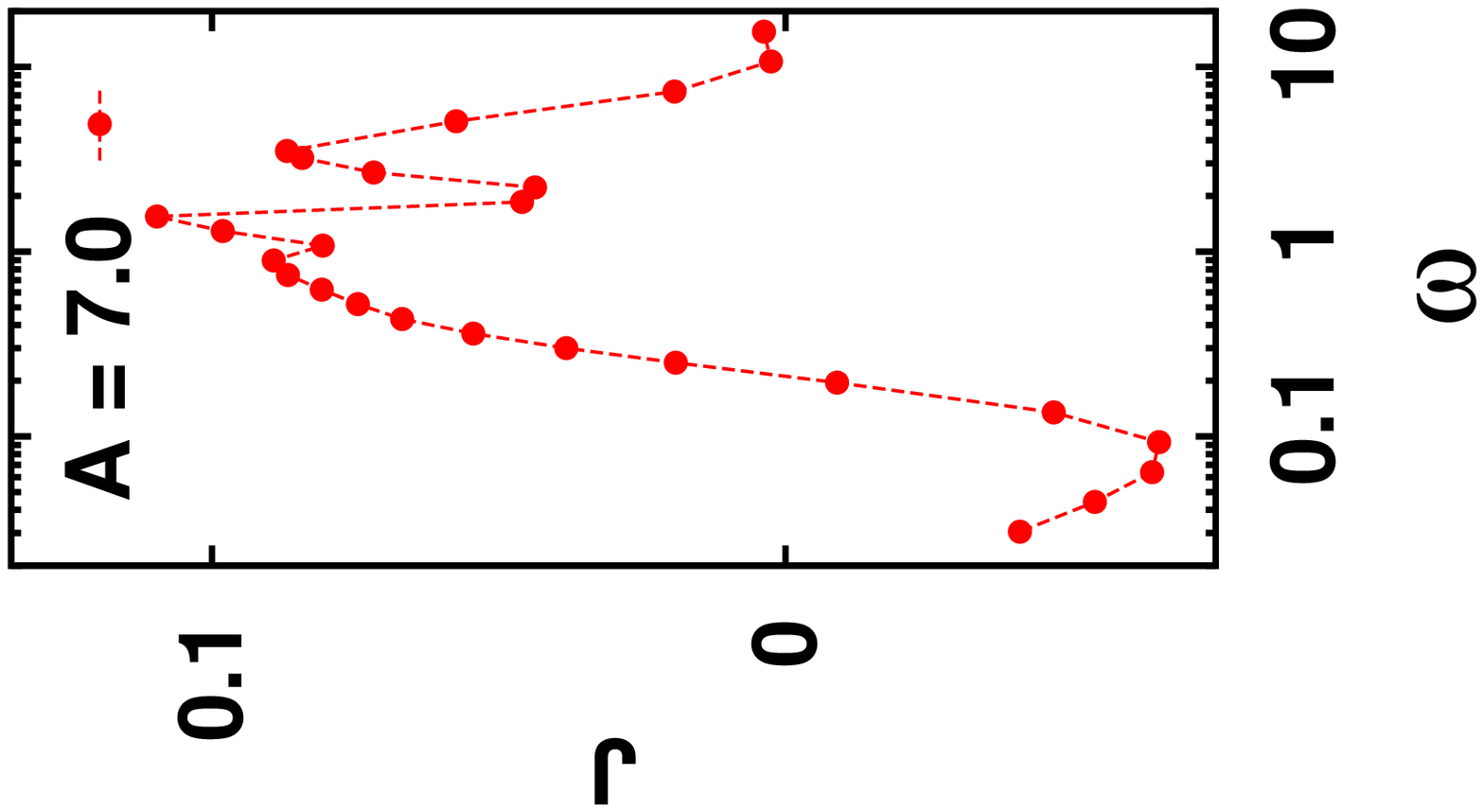}\\
\hskip-1.05cm
\includegraphics[width=5.1cm,angle=-90]{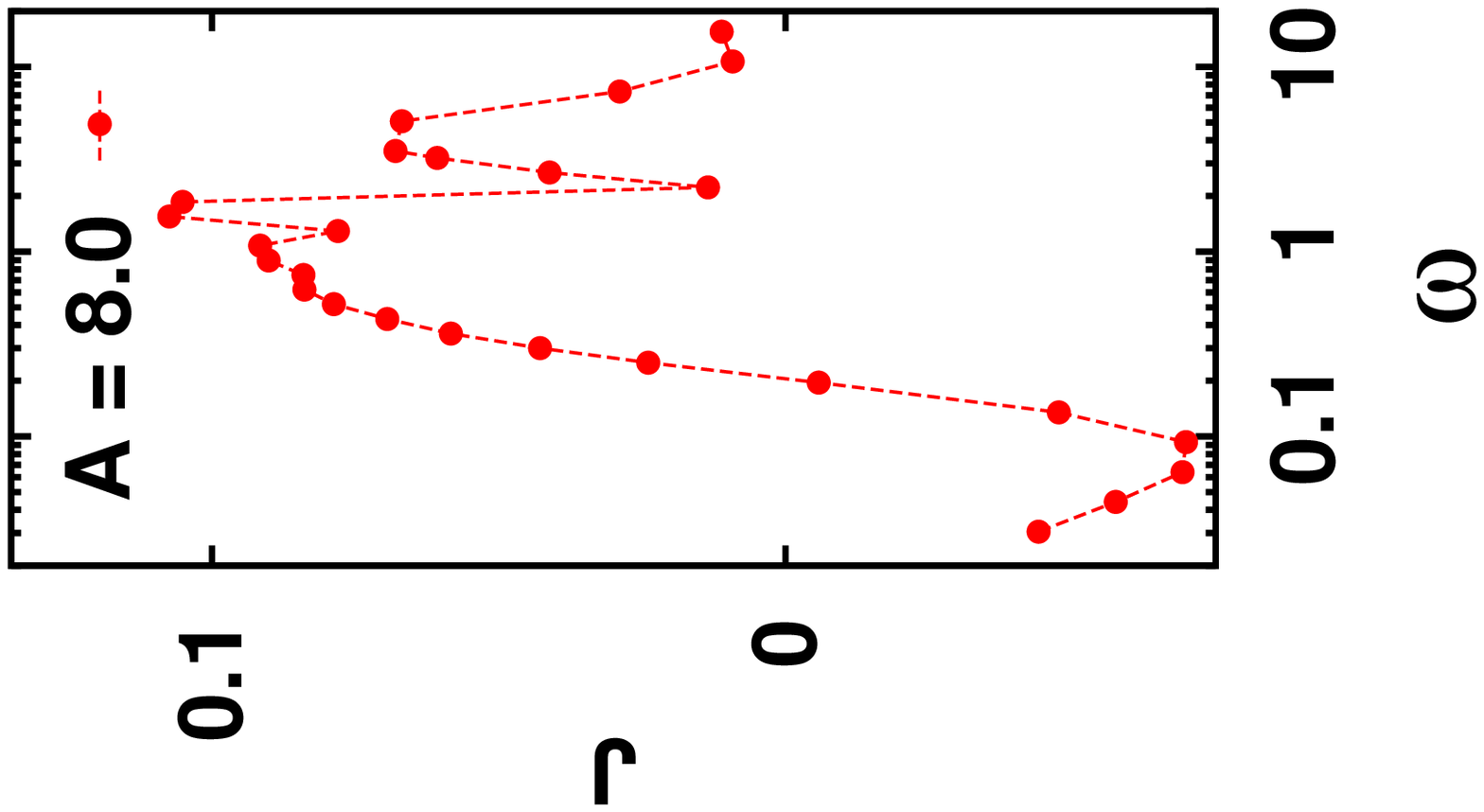}\hskip-0.8cm
\includegraphics[width=5.1cm,angle=-90]{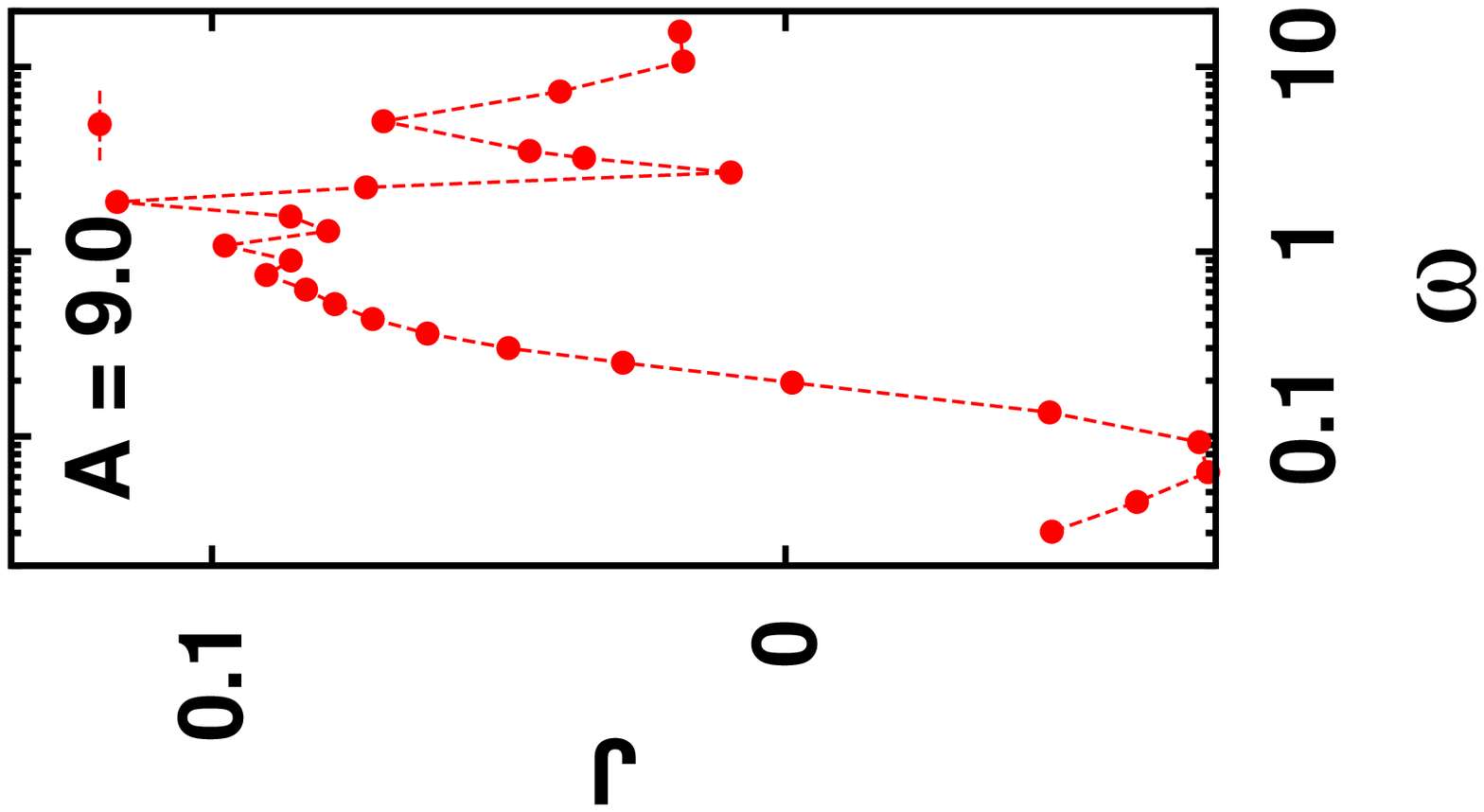}\hskip-0.8cm
\includegraphics[width=5.1cm,angle=-90]{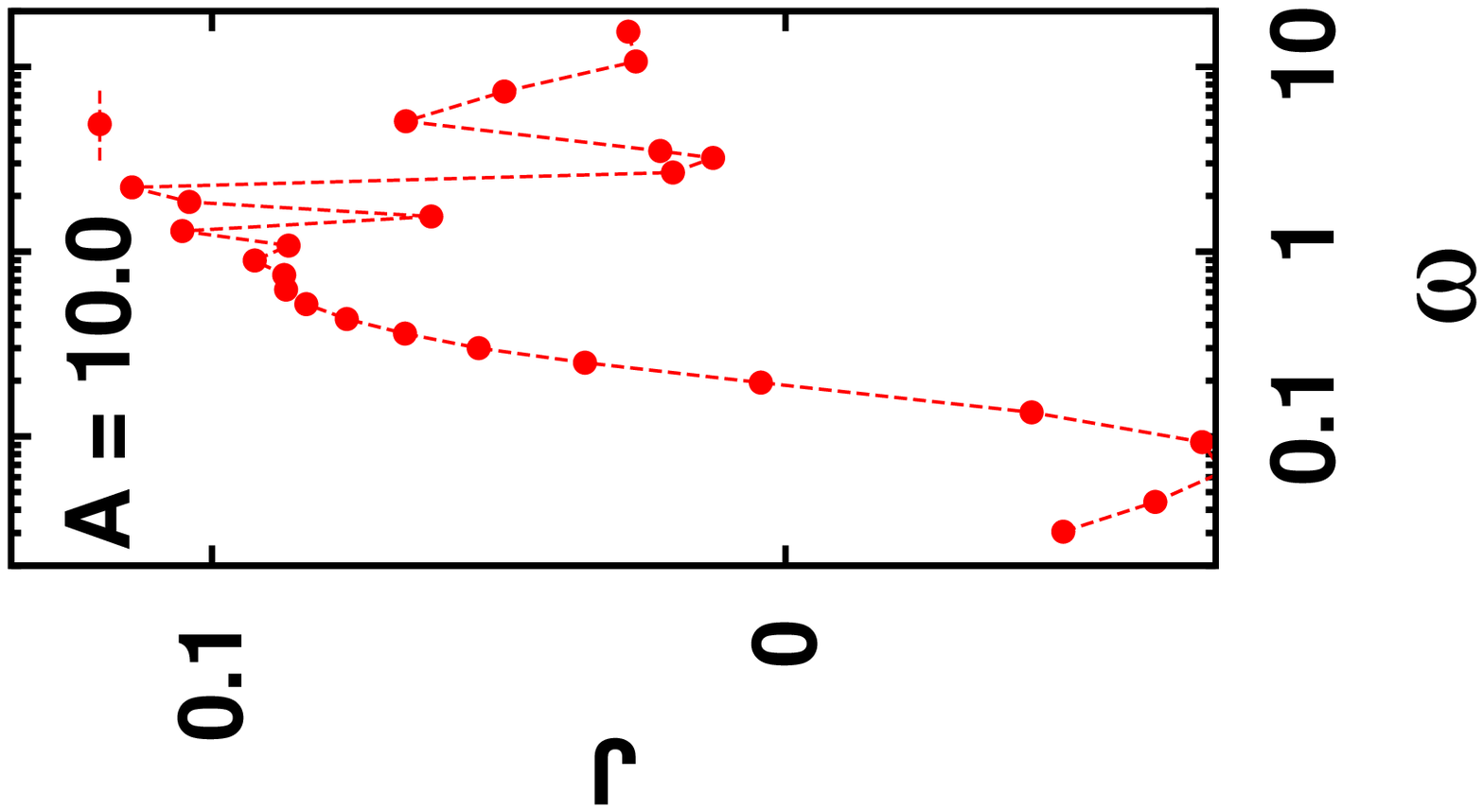}
\caption{(Color online)
The thermal current $J$ for forcing amplitude
$A = 5.0, 6.0,7.0, 8.0, 9.0$ and $10.0$, for bath energy $E_b = -0.1$ with $\Delta E = 0.015$
in a system size of $N = 100$  shows more than one resonance peak as the
forcing amplitude $A$ is increased.
}
\label{fig:mreso_A}
\end{figure}

For certain parameter ranges, the thermal resonance described above can show multiresonance
behavior in which the resonance peak splits into two or more distinct resonance peaks.
In the following, we study the multiresonance phenomenon and its dependence on the
system parameters. In Fig. \ref{fig:mreso_A} the thermal current $J$ is shown as a function of
the forcing frequency $\omega$ for several values of the forcing amplitude $A$. We find that as $A$
increases, first, the central peak splits up in two peaks and the current $J$ corresponding to 
$\omega = \omega_m$ is no longer the maximum now; two other peaks appear on both sides of 
$\omega_m$. Thus in Fig. \ref{fig:mreso_A} we find that for $A=5.0$ there are two peaks, 
whereas, for $A=7.0$ there are three peaks, and for $A = 9.0$ four peaks appear
which are further amplified as $A$ increases. This process continues and new peaks emerge
as the forcing amplitude is increased. Thus multiresonance is enhanced as the amplitude of
the forcing is increased.

\begin{figure}[htbp]
\includegraphics[width=6.25cm,angle=-90]{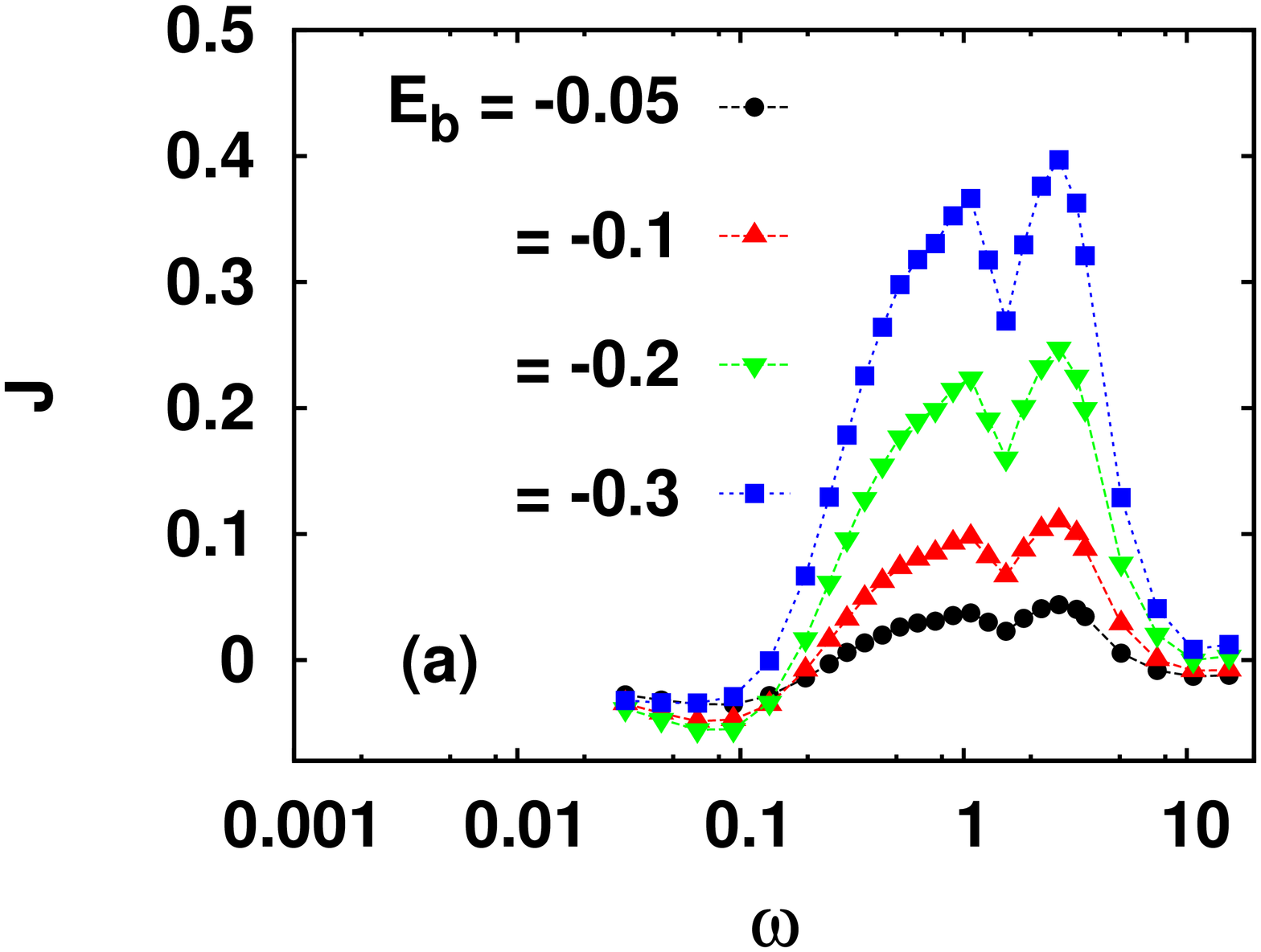}\\
\includegraphics[width=6.25cm,angle=-90]{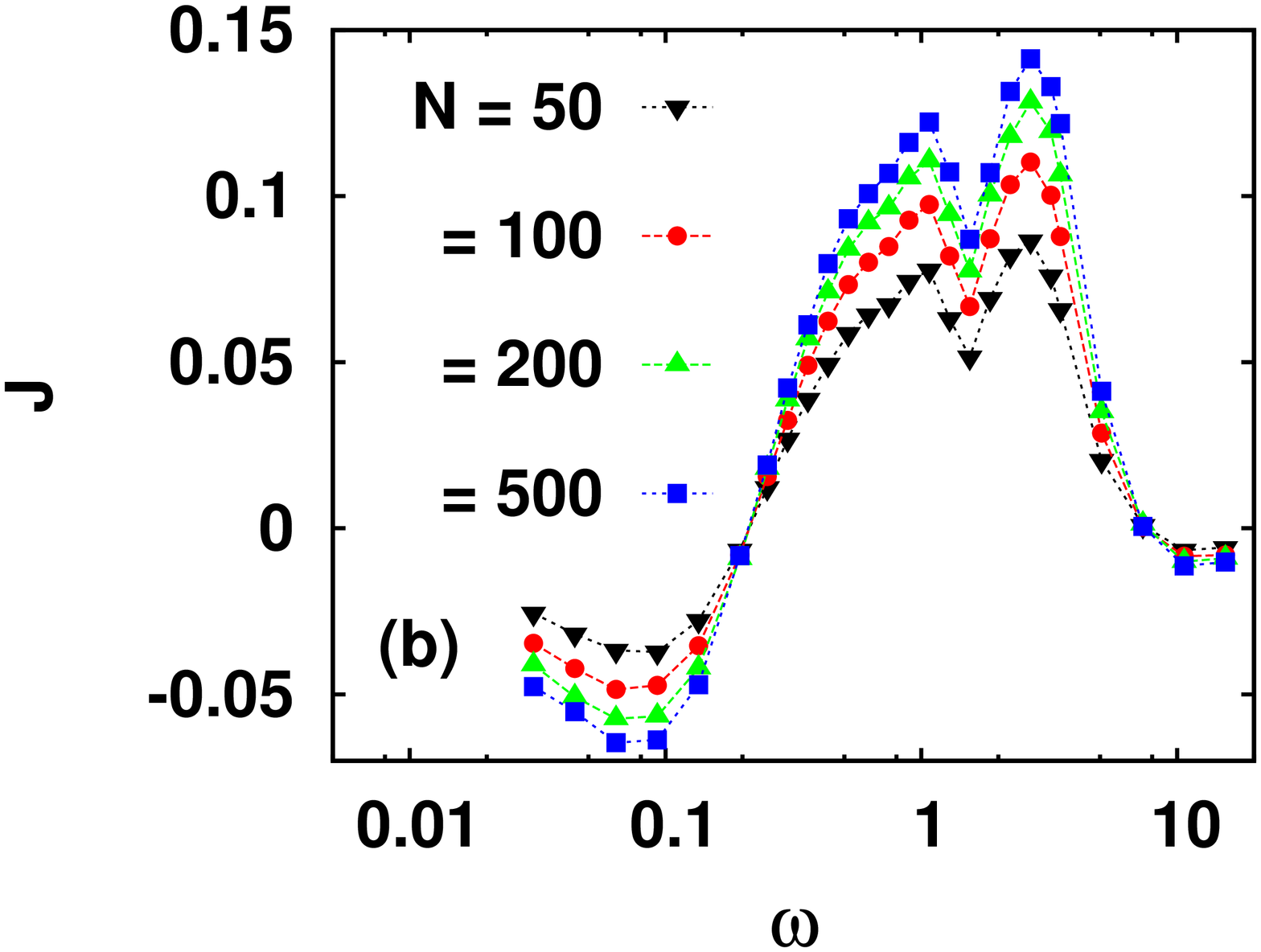}
\caption{(Color online) 
(a) The thermal current $J$ for different average bath energy
$E_b = -0.05, -0.1, -0.2$ and $-0.3$, with $\Delta E = 0.015$ and forcing amplitude $A = 5.0$
for a system size of $N = 100$.
(b) The thermal current $J$ for different system sizes
$N = 50, 100, 200$ and $500$ with $A = 5.0$, $E_b = -0.1$ and with $\Delta E = 0.015$.
}
\label{fig:mreso_TL}
\end{figure}

We also find that the multiresonance feature gets magnified as the average temperature
of the system goes down, similar to the previous case of single peak resonance.
For the FK model, multiresonace is seen both in the limit of low and high temperatures
since the model reduces to an effective harmonic model in both the limits, whereas, 
in the intermediate temperature range there occurs a crossover from single peak to
multiresonace \cite{multi}. In Fig. \ref{fig:mreso_TL}a we show the
multiresonace in our model for different values of the average bath energy $E_b$.
This enhancement in multiresonance can be explained by considering the fact that
the classical Heisenberg model
at low temperatures can be effectively thought to be a harmonic-like system \cite{our}
in terms of the relevant degrees of freedom (here, the angle between the spin vectors $\theta_i$).
This is similar to the case of FK model, where with the increase (or decrease) of temperature a
harmonic-like behavior emerges. Since resonant magnitudes are larger in the harmonic
lattice as compared to the FK model \cite{multi}, we expect an enhanced multiresonance
in the Heisenberg model at lower temperatures. However, the number of resonance peaks seem
to be independent of the average temperature of the system.

The multiresonace feature of our model, however, appears to be different from that
of the FK/harmonic lattices in the sense that only a few spin modes are excited as
the forcing amplitude is increased. It is also observed (from the data shown in
Fig. \ref{fig:mreso_A}) that a modal frequency which is a dominant
mode of energy transport for a given forcing amplitude does not always 
remain a dominant mode when the amplitude is altered.
For example, the dominant frequency for $A = 1.0$ which approximately corresponds to
$\omega_m = 1.5$ is no longer a dominant mode for $A = 5.0$;
two other frequencies on either side of $\omega_m$ become the dominant
modal frequencies that contribute the most to the thermal transport.
Unlike the harmonic system where the amplitude of the forcing signal excites 
all the eigenmodes of the system, here, external forcing selectively
picks up certain frequencies and amplifies them. 
This disparity is also apparent from the fact that the number of modes that are
excited by the external forcing seems to be independent of the system size as can
be seen in Fig. \ref{fig:mreso_TL}b. For the harmonic system, the number of modes
participating in transport is directly proportional to the size of the system which
is definitely not the case in our model.

We have also increased the data sampling frequency to resolve finer structures, if present,
within the broad multiresonace peaks but could not find any, even for very small systems 
at low temperatures for which the harmonic approximation should be valid.
Thus we speculate that, although nonlinearty effects are relatively suppressed
at low temperatures (where our model has an effective harmonic description), 
they do not vanish completely and spin modes remain strongly coupled to each other.
Unlike the FK model, increasing the forcing amplitude to quite larger values in
this model can excite only a few spin modes because of the existence of strong
intrinsic nonlinearity. A more detailed study of the classical Heisenberg model
is surely desirable to unravel the underlying physics of its multiresonance feature.

\subsection{Absence of thermal pumping}
Although the system in resonance exhibits a reversed bulk current, there is no
thermal pumping in this system. This is consistent with the result (previously
obtained for FK/harmonic lattices) that there is no thermal pumping in
force-driven lattices. For a system to be a thermal pump, energy must
be pumped from the colder heat bath and absorbed in the hotter heat bath and
as a consequence a reverse flow of energy occurs in the bulk of the system.
However in our system (and other force-driven lattices in general), although the
high temperature bath absorbs energy from the system and a reversed current flows
through the bulk, the low temperature heat bath does not pump energy into the bulk
of the system. What actually happens is that the additional energy, drawn
from the external periodic forcing, flows from the point of forcing (here $i = 1$)
towards the two boundaries of the system, and thus, results in a reversed bulk current.

\begin{figure}[htbp]
\hskip-0.85cm
\includegraphics[width=6cm,angle=-90]{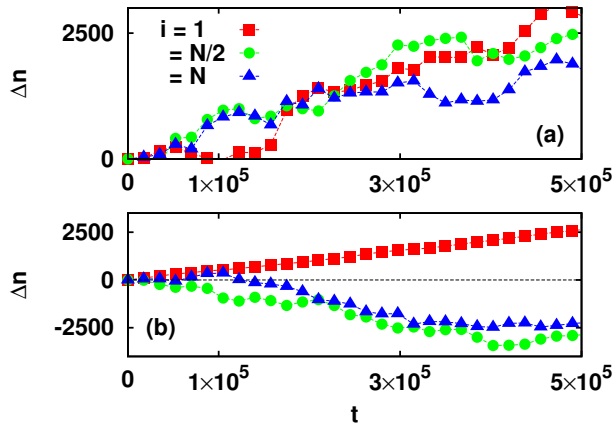}
\caption{(Color online) Computation of $\Delta n = n_l - n_r$ from a typical
run using the DTOE dynamics for three sites $i = 1, N/2$ and $N$ in a system of
size $N = 100$, $E_b = -0.1$, $\Delta E = 0.015$:
(a) without periodic forcing energy across all the sites flow towards left;
(b) with periodic forcing ($A = 1.0$, $\omega = 1.5$) energy across site $i = 1$ flow towards left
whereas, for sites $i = N/2$ and $N$ flows to the right. Thus both the heat 
baths absorb energy and there is no thermal pumping. 
}
\label{fig:flow}
\end{figure}

This can be clearly seen in our system by monitoring the energy flow across sites
$i = 1, N/2$ and $N$. When the spin at a site $i$
is updated using the DTOE dynamics, the energy of the bonds connected to it,
namely $E_{i-1}$ and $E_i$, are updated such that the sum $E_{i-1} + E_i$ remains the
same before and after the update. In other words, a redistribution of the
sum $E_{i-1}+E_i$ takes place while the $i$-th spin is updated. Since our
system is connected to stochastic thermal baths, this redistribution process is
also stochastic in nature - the $(i-1)$-th bond stochastically gains and
loses energy and similarly for the $i$-th bond. However, since there is an overall
thermal gradient $(T_r - T_l)$, the total number of times energy flows in one
particular direction (here, towards left since $ T_r > T_l$) over a long time will
obviously be more than that of the opposite direction so that a steady energy
current flows through the system.
In absence of the periodic forcing, the quantity $\Delta n = n_l - n_r$ is positive
when measured for sites $i = 1, N/2$ and $N$, where $n_l$ ($n_r$) is the total
number of times energy flows to the left (right) across a particular site. 
The results from a typical run, starting from a random
initial condition of spins and using the DTOE dynamics, is shown in Fig.
\ref{fig:flow}a. Thus energy flows from the high energy bath to the low
energy bath i.e., from right to left end of the system.
However when the system is in resonance region, $\Delta n$ for site $i = 1$ is 
positive whereas that for  sites $i = N/2$ and $N$ is negative. This shows that
while in resonance, energy flows from right to left for  $i = 1$ but for $i = N/2$
and the right end $i=N$, energy flows from left to right (see Fig. \ref{fig:flow}b). 
Thus current flows towards the two boundaries from the point of periodic perturbation
in the bulk of the system. 
Evidently, there is no thermal pumping since both the high temperature and low temperature
baths absorb energy and thus the transport of energy from low to high temperature bath
is absent.

\section{Summary}
\label{summary}
To summarise, we report here the extensive numerical study of the one dimensional
classical Heisenberg model in presence of boundary drive and time varying forcing. 
A time-periodic magnetic field acts locally at one end of the system (site $i = 1$)
and a thermal gradient is maintained by boundary heat baths. We choose the time-periodic
forcing to be sinusoidal with amplitude $A$ and frequency $\omega$. The thermal current
that flows though the system shows resonance at some characteristic frequency
$\omega_m$ of the forcing frequency for which the current attains a maximum value.
By properly tuning the boundary temperatures,  we demonstrate that the energy
current flowing through the bulk of the system can be reversed for frequencies
within the resonance region. The magnitude of the current can also be controlled
by tuning the forcing amplitude $A$. This allows one to mechanically control the
magnitude as well as the direction of current in the system. This could be of immense
practical use in nanoscale systems where a large thermal gradient is not desirable.

We study the dependence of the thermal resonance on the parameters of the system.
It is found that the magnitude of resonance increases as the system size increases and
survives in the thermodynamic limit. The maximum thermal current $J_m$ corresponding
to the resonance frequency $\omega_m$, saturates to a finite value for a
thermodynamically large system. Also decreasing the average temperature enhances
the magnitude of resonance; the maximum thermal current $J_m$ varies as $T^{-1}$
at low temperatures in the thermodynamic limit.

In some parameter range, the single resonance peak splits into multiple peaks and a
multiresonance phenomenon is observed. As the amplitude of the external forcing
is increased, the number of peaks also increases. The resonance and the
multiresonace phenomenon can in general be explained as follows. 
The external periodic forcing that is imposed on the system acts as an
additional source of energy for the system. When the frequency of
the forcing coincides with the natural frequencies of the system, the transfer
of energy from the external perturbation to the system becomes maximum. 
However for our system, certain frequencies are selectively excited by the external
forcing and the the number of modal frequencies that participate in thermal transport
is determined by the forcing amplitude. Similar to single peak resonance, the magnitude
of multiresonance is also enhanced with the decrease of average temperature of the
system; the modal frequencies and their number, however, seem to be independent
of the average temperature and also the size of the system.

Finally, we explicitly show using energy flow arguments that despite the reversal of the
current in the bulk, this system fails to act as a thermal pump. This is consistent
with the previous result that a force-driven lattice can not direct thermal energy
from the low temperature heat bath to the high temperature heat bath.

The classical model can be thought to be the infinitely large spin limit of
the quantum Heisenberg model. This classical approximation of the quantum model 
is already seen to hold for systems with spin $s = 5/2$, for example, in $Mn^{2+}$ \cite{windsor}.
As such, controlled laboratory experiments with model chemical compounds, 
which are now-a-days routinely performed, can be used to test the theoretical
predictions made in the classical Heisenberg model. Hopefully, such experimental studies
will eventually lead us towards better heat control and management in future.

\textbf{Acknowledgement:} The author would like to thank P. K. Mohanty for 
 helpful suggestions and careful reading of the manuscript.


\end{document}